\documentclass[12pt,a4paper]{article}
\usepackage{amssymb}
\usepackage{graphicx}

\def\1{{1\mskip-10mu1}}

\def\bea{\begin{eqnarray*}}
\def\eea{\end{eqnarray*}}
\def\bean{\begin{eqnarray}}
\def\eean{\end{eqnarray}}

\newtheorem{ftheo}{THEOREM}[section]

\newtheorem{flemma}[ftheo]{LEMMA}

\begin{document}

\title{Semiclassical Properties and Chaos Degree for the Quantum Baker's Map}
\author{Kei Inoue$\dagger $ , Masanori Ohya$\dagger $ and Igor V. Volovich$\ddagger $
\\
$\dagger $Department of Information Sciences\\
Science University of Tokyo\\
Noda City, Chiba 278-8510 Japan\\
$\ddagger ~$Steklov Mathematical Institute,\\
Russian Academy of Science\\
Gubkin St. 8, Moscow, GSP1, 117966,\\
Russia\\
email:volovich@mi.ras.ru}
\maketitle
\date{}

\begin{abstract}
We study the chaotic behaviour and the quantum-classical correspondence for
the baker's map. Correspondence between quantum and classical expectation
values is investigated and it is numerically shown that it is lost at the
logarithmic timescale. The quantum chaos degree is computed and it is
demonstrated that it describes the chaotic features of the model. The
correspondence between classical and quantum chaos degrees is considered.
\end{abstract}

\section{Introduction}

The study of chaotic behaviour in classical dynamical systems is dating back
to Lobachevsky and Hadamard who have been studied the exponential
instability property of geodesics on manifolds of negative curvature and to
Poincare, who initiated the inquiry into the stability of the solar system.
One believes now that the main features of chaotic behaviour in the
classical dynamical systems are rather well understood, see for example \cite
{AA, Sin}. However the status of ``quantum chaos'' is much less clear
although the significant progress has been made on this front.

Sometimes one says that an approach to quantum chaos, which attempts to
generalize the classical notion of sensitivity to initial conditions, fails
for two reasons: first there is no quantum analogue of the classical phase
space trajectories and, second, the unitarity of linear Schrodinger equation
precludes sensitivity to initial conditions in the quantum dynamics of state
vector. Let us remind, however, that in fact there exists a quantum analogue
of the classical phase space trajectories. It is quantum evolution of
expectation values of appropriate observables in suitable states. Also let
us remind that the dynamics of a classical system can be described either by
the Hamilton equations or by the liner Liouville equations. In quantum
theory the linear Schrodinger equation is the counterpart of the Liouville
equation while the quantum counterpart of the classical Hamilton's equation
is the Heisenberg equation. Therefore the study of quantum expectation
values should reveal the chaotic behaviour of quantum systems. In this paper
we demonstrate this fact for the quantum baker's map.

If one has the classical Hamilton's equations

\[
dq/dt=p,\quad dp/dt=-V^{\prime }\left( q\right) , 
\]

\noindent then the corresponding quantum Heisenberg equations have the same
form

\[
dq_{h}/dt=p_{h},\quad dp_{h}/dt=-V^{\prime }\left( q_{h}\right) , 
\]

\noindent where $q_{h}$ and $p_{h}$ are quantum canonical operators of
position and momentum. For the expectation values one gets the Ehrenfest
equations

\[
d<q_{h}>/dt=<p_{h}>,\quad d<p_{h}>/dt=-<V^{\prime }\left( q_{h}\right) > 
\]

Note that the Ehrenfest equations are classical equations but for nonlinear $%
V^{\prime }\left( q_{h}\right) $ they are neither Hamiltonian's equations
nor even differential equations because one can not write $<V^{\prime
}\left( q_{h}\right) >$ as a function of $<q_{h}>$ and $<p_{h}>.$ However
these equations are very convenient for the consideration of the
semiclassical properties of quantum system. The expectation values $<q_{h}>$
and $<p_{h}>$ are functions of time and initial data. They also depend on
the quantum states. One of important problems is to study the dependence of
expectation values from the initial data. In this paper we will study this
problem for the quantum baker's map.

The main objective of ``quantum chaos'' is to study the correspondence
between classical chaotic systems and their quantum counterparts in the
semiclassical limit \cite{Gut,CC}$.\smallskip $ The quantum-classical
correspondence for dynamical systems has been studied for many years, see
for example\cite{AMRV,Hep,Zur, Zur2,ENTS,Has} and reference therein. A
significant progress in understanding of this correspondence has been
achieved in the WKB approach when one considers the Planck constant $h$ as a
small variable parameter. Then it is well known that in the limit $%
h\rightarrow 0$ quantum theory is reduced to the classical one \cite{Mas}.
However in physics the Planck constant is a fixed constant although it is
very small. Therefore it is important to study the relation between
classical and quantum evolutions when the Planck constant is fixed. There is
a conjecture \cite{BZ2,Zas,Ber,Zur2} that a characteristic timescale $\tau $
appears in the quantal evolution of chaotic dynamical systems. For time less
then $\tau $ there is a correspondence between quantum and classical
expectation values, while for times greater that $\tau $ the predictions of
the classical and quantum dynamics no longer coincide. The important problem
is to estimate the dependence $\tau $ on the Planck constant $h.$ Probably a
universal formula expressing $\tau $ in terms of $h$ does not exist and
every model should be studied case by case. It is expected that certain
quantum and classical expectation values diverge on a timescale inversely
proportional to some power of $h$ \cite{BV}. Other authors suggest that a
breakdown may be anticipated on a much smaller logarithmic timescale \cite
{OS,DDS,SV,KH,LO,SC1,SC2,OT}. The characteristic time $\tau $ associated
with the hyperbolic fixed points of the classical motion is expected to be
of the logarithmic form $\tau =\frac{1}{\lambda }\ln \frac{C}{h}$ where $%
\lambda $ is the Lyapunov exponent and $C$ is a constant which can be taken
to be the classical action. Such the logarithmic timescale has been found in
the numerical simulations of some dynamical models.\cite{Zur}. It was shown
also that the discrepancy between quantum and classical evolutions is
decreased by even a small coupling with the environment, which in the
quantum case leads to decoherence \cite{Zur}.

The chaotic behaviour of the classical dynamical systems is often
investigated by computing the Lyapunov exponents. An alternative quantity
measuring chaos in dynamical systems which is called the chaos degree has
been suggested in \cite{O1} in the general framework of information dynamics 
\cite{IKO}. The chaos degree was applied to various models in \cite{IOS}. An
advantage of the chaos degree is that it can be applied not only to
classical systems but also to quantum systems as well.

In this work we study the chaotic behaviour and the quantum-classical
correspondence for the baker's map \cite{BV,Sar}. The quantum baker's map is
a simple model invented for the theoretical study of quantum chaos. Its
mathematical properties have been studied in numerical works. In particular
its semiclassical properties have been considered \cite
{OS,DDS,SV,KH,LO,SC1,SC2,OT}, quantum computing and optical realizations
have been proposed \cite{HKO,Sch,BS}, various quantization procedures have
been discussed \cite{LB,Lak,SV,SC3}, a symbolic dynamics representation has
been given \cite{SC3}.

It is well known that for the consideration of the semiclassical limit in
quantum mechanics it is very useful to use coherent states. We define an
analogue of the coherent states for the quantum baker's map. We study the
quantum baker's map by using the correlation functions of the special form
which corresponds to the expectation values of Weyl operators, translated in
time by the unitary evolution operator and taken in the coherent states.

To explain our formalism we first discuss the classical limit for
correlation functions in ordinary quantum mechanics. Correspondence between
quantum and classical expectation values for the baker's map is investigated
and it is numerically shown that it is lost at the logarithmic timescale.
The chaos degree for the quantum baker's map is computed and it is
demonstrated that it describes the chaotic features of the model. The
dependence of the chaos degree on the Planck constant is studied and the
correspondence between classical and quantum chaos degrees is established.

\section{ Quantum vs. Classical Dynamics}

In this section we discuss an approach to the semiclassical limit in quantum
mechanics by using the coherent states, see \cite{Hep}. Then in the next
section an extension of this approach to the quantum baker's map will be
given.

\noindent Consider the canonical system with the Hamiltonian function

\begin{equation}
H =\frac{p^{2}}{2}+V\left( x\right)  \label{2.1}
\end{equation}
in the plane $\left( p,x\right) \in \mathbf{R}^{2}$ . We assume that the
canonical equations

\begin{equation}
\dot{x}\left( t\right) =p\left( t\right) ,\mbox{\hspace{1cm}}\dot{p}\left(
t\right) =-V^{\prime }\left( x\left( t\right) \right)  \label{2.2}
\end{equation}
have a unique solution $\left( x\left( t\right) ,p\left( t\right) \right) $
for times $|t|<T$ with the initial data

\begin{equation}
x\left( 0\right) =x_{0},\hspace{1cm}p\left( 0\right) =v_{0}  \label{2.3}
\end{equation}
This is equivalent to the solution of the Newton equation

\begin{equation}
\ddot{x}\left( t\right) =-V^{\prime }\left( x\left( t\right) \right)
\label{2.4}
\end{equation}
with the initial data

\begin{equation}
x\left( 0\right) =x_{0},\hspace{1cm}\dot{x}\left( 0\right) =v_{0}
\label{2.5}
\end{equation}
We denote

\begin{equation}
\alpha =\frac{1}{\sqrt{2}}\left( x_{0}+iv_{0}\right)  \label{2.6}
\end{equation}

\noindent The quantum Hamiltonian operator has the form

\[
H_{h}=\frac{p_{h}^{2}}{2}+V\left( q_{h}\right) 
\]
where $p_{h}$ and $q_{h}$ satisfy the commutation relations

\[
\left[ p_{h},q_{h}\right] =-ih 
\]
The Heisenberg evolution of the canonical variables is defined as

\[
p_{h}\left( t\right) =U\left( t\right) p_{h}U\left( t\right) ^{*}, %
\mbox{\hspace{1cm}}q_{h}\left( t\right) =U\left( t\right) q_{h}U\left(
t\right) ^{*} 
\]
where

\[
U\left( t\right) =\exp \left( -itH_{h}/h\right) 
\]
For the consideration of the classical limit we take the following
representation

\[
p_{h}=-ih^{1/2}\partial /\partial x,\mbox{\hspace{1cm}}q_{h}=h^{1/2}x 
\]
acting to functions of the variable $x$ $\in $ $\mathbf{R.}$ We also set

\[
a=\frac{1}{\sqrt{2}h^{1/2}}\left( q_{h}+ip_{h}\right) =\frac{1}{\sqrt{2}}
\left( x+\frac{\partial }{\partial x}\right) ,\;a^{*}=\frac{1}{\sqrt{2}%
h^{1/2}}\left( q_{h}-ip_{h}\right) =\frac{1}{\sqrt{2}}\left( x-\frac{%
\partial }{\partial x}\right) , 
\]
then

\[
\left[ a,a^{*}\right] =1. 
\]
The coherent state $\left| \alpha \right\rangle $ is defined as

\begin{equation}
\left| \alpha \right\rangle =W\left( \alpha \right) \left| 0\right\rangle
\label{2.7}
\end{equation}
where $\alpha $ is a complex number, $W\left( \alpha \right) =\exp
\left(\alpha a^{*}-a\alpha ^{*}\right) $ and $\left| 0\right\rangle $ is the
vacuum vector, $a\left| 0\right\rangle =0.$ The vacuum vector is the
solution of the equation

\begin{equation}
\left( q_{h}+ip_{h}\right) \left| 0\right\rangle =0  \label{2.8}
\end{equation}
In the $x$ - representation one has

\begin{equation}
\left| 0\right\rangle =\exp \left( -x^{2}/2\right) /\sqrt{2\pi }.
\label{2.9}
\end{equation}
The operator $W\left( \alpha \right) $ one can write also in the form

\begin{equation}
W\left( \alpha \right) =Ce^{iq_{h}v_{0}/h^{1/2}}e^{-ip_{h}x_{0}/h^{1/2}}
\label{2.10}
\end{equation}
where $C=\exp \left( -v_{0}x_{0}/2h\right) .$

The mean value of the position operator with respect to the coherent vectors
is the real valued function

\begin{equation}
q\left( t,\alpha ,h\right) =\left\langle h^{-1/2}\alpha \right| q_{h}\left(
t\right) \left| h^{-1/2}\alpha \right\rangle  \label{2.11}
\end{equation}
Now one can present the following basic formula describing the semiclassical
limit

\begin{equation}
\lim_{h\rightarrow 0}q\left( t,\alpha ,h\right) =x\left( t,\alpha \right)
\label{2.12}
\end{equation}
Here $x\left( t,\alpha \right) $ is the solution of (\ref{2.4}) with the
initial data (\ref{2.5}) and $\alpha $ is given by (\ref{2.6}).

Let us notice that for time $t=0$ the quantum expectation value $q\left(
t,\alpha ,h\right) $ is equal to the classical one:

\begin{equation}
q\left( 0,\alpha ,h\right) =x\left( 0,\alpha \right) =x_{0}  \label{2.13}
\end{equation}
for any $h.$ We are going to compare the time dependence of two real
functions $q\left( t,\alpha ,h\right) $ and $x\left( t,\alpha \right) $ 
these functions are approximately equal. The important problem is to
estimate for which $t$ the large difference between them will appear. It is
expected that certain quantum and classical expectation values diverge on a
timescale inversely proportional to some power of $h$ \cite{Hep}. Other
authors suggest that a breakdown may be anticipated on a much smaller
logarithmic timescale \cite{OS,DDS,SV,KH,LO,SC1,SC2,OT}. One of very
interesting examples \cite{AMRV} of classical systems with chaotic behaviour
is described by the hamiltonian function

\[
H =\frac{p_{1}^{2}}{2}+\frac{p_{2}^{2}}{2}+\lambda x_{1}^{2}x_{2}^{2} 
\]
The consideration of this classical and quantum model within the described
framework will be presented in another publication.

\section{Coherent States for the Quantum Baker's Map}

The classical baker's transformation maps the unit square $0\leq $ $q,p\leq
1 $ onto itself according to

\[
\left( q,p\right) \rightarrow \left\{ 
\begin{array}{ll}
\left( 2q,p/2\right) , & \mbox{if}\quad 0\leq q\leq 1/2 \\ 
\left( 2q-1,\left( p+1\right) /2\right) , & \mbox{if\quad }1/2<q\leq 1
\end{array}
\right. 
\]
This corresponds to compressing the unit square in the $p$ direction and
stretching it in the $q$ direction, while preserving the area, then cutting
it vertically and stacking the right part on top of the left part.

The classical baker's map has a simple description in terms of its symbolic
dynamics \cite{AY}. Each point $\left( q,p\right) $ is represented by a
symbolic string

\begin{equation}
\xi =\cdots \xi _{\_2}\xi _{\_1}\xi _{0}.\xi _{1}\xi _{2}\cdots ,
\label{3.1}
\end{equation}

\noindent where $\xi _{k}\in \left\{ 0,1\right\} $, and

\[
q=\sum_{k=1}^{\infty }\xi _{k}2^{-k},\qquad p=\sum_{k=0}^{\infty }\xi
_{-k}2^{-k-1} 
\]

\noindent The action of the baker's map on a symbolic string $s$ is given by
the shift map (Bernoulli shift ) $U$ defined by $U\xi =\xi ^{^{\prime }}$,
where $\xi _{k}^{^{\prime }}=\xi _{k+1}$. This means that, at each time
step, the dot is shifted one place to the right while entire string remains
fixed. After $n$ steps the $q$ coordinate becomes

\begin{equation}
q_{n}=\sum_{k=1}^{\infty }\xi _{n+k}2^{-k}  \label{3.2}
\end{equation}
This relation defines the classical trajectory with the initial data

\begin{equation}
q=q_{0}=\sum_{k=1}^{\infty }\xi _{k}2^{-k}  \label{3.3}
\end{equation}

Quantum baker's maps are defined on the $D$-dimensional Hilbert space of the
quantized unit square. To quantize the unite square one defines the Weyl
unitary displacement operators $\hat{U}$ and $\hat{V}$ in $D$ - dimensional
Hilbert space, which produces displacements in the momentum and position
directions, respectively, and the following commutation relation is obeyed

\[
\hat{U}\hat{V}=\epsilon \hat{V}\hat{U}, 
\]

\noindent where $\epsilon =\exp \left( 2\pi i/D\right) .$ We choose $D=2^{N}$
, so that our Hilbert space will be the $N$ qubit space $\Bbb{C}^{\otimes N}$%
. The constant $h=1/D=2^{-N}$ can be regarded as the Plank constant. The
space $\Bbb{C}^{2}$ has a basis

\[
\left| 0\right\rangle =\left( 
\begin{array}{c}
0 \\ 
1
\end{array}
\right) ,\left| 1\right\rangle =\left( 
\begin{array}{c}
1 \\ 
0
\end{array}
\right) 
\]
The basis in $\Bbb{C}^{\otimes N}$ is

\[
\left| \xi _{1}\right\rangle \otimes \left| \xi _{2}\right\rangle \otimes
\cdots \otimes \left| \xi _{N}\right\rangle ,~~\xi _{k}=0,1 
\]
We write

\[
\xi =\sum_{k=1}^{N}\xi _{k}2^{N-k} 
\]
then $\xi =0,1,...,2^{N}-1$ and denote

\[
\left| \xi \right\rangle =\left| \xi _{1}\xi _{2}\cdots \xi_{N}\right\rangle
=\left| \xi _{1}\right\rangle \otimes \left| \xi_{2}\right\rangle \otimes
\cdots \otimes \left| \xi _{N}\right\rangle 
\]
We will use for this basis also notations $\{\left| \eta \right\rangle
=\left| \eta _{1}\eta _{2}\cdots \eta _{N}\right\rangle ,~~\eta _{k}=0,1\}$
and $\{\left| j\right\rangle =\left| j_{1}j_{2}\cdots
j_{N}\right\rangle,j_{k}=0,1\}$.

The operators $\hat{U}$ and $\hat{V}$ can be written as 
\[
\hat{U}=e^{2\pi i\hat{q}},\quad \hat{V}=e^{2\pi i\hat{p}} 
\]
where the position and momentum operators $\hat{q}$ and $\hat{p}$ are
operators in $\Bbb{C}^{\otimes N}$ which are defined as follows. The
position operator is 
\[
\hat{q}=\sum_{j=0}^{2^{N}-1}q_{j}\left| j\right\rangle \left\langle
j\right|=\sum_{j_{1},...,j_{N}}q_{j}\left| j_{N}...j_{1}\right\rangle
\left\langle j_{1}...j_{N}\right| 
\]
where

\[
\left| j\right\rangle =\left| j_{1}j_{2}\cdots j_{N}\right\rangle ,
j_{k}=0,1 
\]
is the basis in $\Bbb{C}^{\otimes N}$,

\[
j=\sum_{k=1}^{N}j_{k}2^{N-k} 
\]
and

\[
q_{j}=\frac{j+1/2}{2^{N}},~~j=0,1,\ldots ,2^{N}-1 
\]
The momentum operator is defined as

\[
\hat{p}=F_{N}\hat{q}F_{N}^{*} 
\]

where $F_{N}$ is the quantum Fourier transform acting to the basis vectors as

\[
F_{N}\left| j\right\rangle =\frac{1}{\sqrt{D}}\sum_{\xi =0}^{D-1}e^{2\pi
i\xi j/D}\left| \xi \right\rangle , 
\]
here $D=2^{N}$.

The symbolic representation of quantum baker's map $T$ was introduced by
Schack and Caves \cite{SC3} and studied in \cite{SS1,SS2}. Let us explain
the symbolic representation of quantum baker's map as a special case \cite
{SC3}: By applying a partial quantum Fourier transform $G_{m}=\stackrel{m}{%
\overbrace{I\otimes \cdots \otimes I}}\otimes F_{N-m}$ to the position
eigenstates, one obtains the following quantum baker's map $T$:

\[
T\left| _{\bullet }\xi _{1}\cdots \xi _{N}\right\rangle \equiv \left|
\xi_{1\bullet }\xi _{2}\cdots \xi _{N}\right\rangle , 
\]
where

\[
T=G_{N-1}\circ G_{N}^{-1} 
\]
and

\begin{eqnarray*}
\left| \xi _{1}\cdots \xi _{N-m\bullet }\xi _{N-m+1}\cdots
\xi_{N}\right\rangle &\equiv &G_{m}\left| \xi _{N-m+1}\cdots \xi
_{N}\xi_{N-m}\cdots \xi _{1}\right\rangle \\
&=&\left| \xi _{N-m+1}\right\rangle \otimes \cdots \otimes \left|
\xi_{N}\right\rangle \otimes F_{N-m}\left| \xi _{N-m}\right\rangle \otimes
\cdots \otimes \left| \xi _{1}\right\rangle .
\end{eqnarray*}

The quantum baker`s map $T$ is the unitary operator in $\Bbb{C}^{\otimes N}$
with the following matrix elements

\begin{equation}
\left\langle \xi \right| T\left| \eta \right\rangle =\frac{1-i}{2}\exp
\left( \frac{\pi }{2}i\left| \xi _{1}-\eta _{N}\right| \right)
\prod_{k=2}^{N}\delta \left( \xi _{k}-\eta _{k-1}\right) ,  \label{3.5}
\end{equation}
\noindent where $\left| \xi \right\rangle =\left| \xi _{1}\xi _{2}\cdots
\xi_{N}\right\rangle $,~~ $\left| \eta \right\rangle =\left| \eta
_{1}\eta_{2}\cdots \eta _{N}\right\rangle $ and $\delta (x)$ is the
Kronecker symbol, $\delta (0)=1;~\delta (x)=0,x\neq 0.$

We define the coherent states by

\begin{equation}
\left| \alpha \right\rangle =Ce^{2\pi i\hat{q}v}e^{-2\pi i\hat{p}x}\left|
\psi _{0}\right\rangle  \label{3.6}
\end{equation}

Here $\alpha =x+iv,$ $x$ and $v$ are integers, $C$ is the normalization
constant and $\left| \psi _{0}\right\rangle $ is the vacuum vector. This
definition should be compared with \ref{2.10}. The vacuum vector can be
defined as the solution of the equation

\[
\left( q_{h}+ip_{h}\right) \left| \psi _{0}\right\rangle =0 
\]

\noindent (compare with (\ref{2.8})). We will use the simpler definition
which in the position representation is

\[
\left\langle q_{j}\right. \left| \psi _{0}\right\rangle =C\exp
\left(-q_{j}^{2}/2\right) 
\]

\noindent (compare with (\ref{2.9})). Here $C$ is a normalization constant.

\section{Chaos Degree}

Let us review the entropic chaos degree defined in \cite{O1}. This entropic
chaos degree is given by a probability distribution $\varphi $ and a
dynamics (channel) $\Lambda ^{*}$ sending a state to a state; $\varphi
=\sum_{k}p_{k}\delta _{k},$ where $\delta _{k}$ is the delta measure such as 
$\delta _{k}\left(j\right) \equiv \left\{ 
\begin{array}{ll}
1 & \left( k=j\right) \\ 
0 & \left( k\neq j\right)
\end{array}
\right.$. Then the entropic chaos degree is defined as

\begin{equation}
D\left( \varphi ;\Lambda ^{*}\right) =\sum_{k}p_{k}S(\Lambda ^{*}\delta _{k})
\label{4.1}
\end{equation}

\noindent with the von Neumann entropy $S$, equivalently to the Shannon
entropy because the probability distribution $\varphi $ is a classical
object.

A dynamics $\mathcal{F}$ of the orbit produces the above channel $%
\Lambda^{*} $, so that let $\left\{ x_{n}\right\} $ be the orbit and $%
\mathcal{F}$ be a map from $x_{n}$ to $x_{n+1}$.

Take a finite partition $\left\{ B_{k}\right\} $ of $I=\left[ a,b\right]
^{l} $ $\left( a,b\in \mathbf{R}\right) \subset \mathbf{R}^{l}$such as

\[
I=\bigcup_{k}B_{k}\quad \left( B_{i}\bigcap B_{j}=\emptyset ,i\neq j\right) 
\]

\noindent for a map $\mathcal{F}$ on $I$ with $x_{n+1}=\mathcal{F}\left(
x_{n}\right) $ (a difference equation). The state $\varphi ^{\left( n\right)
}$ of the orbit determined by the difference equation is defined by the
probability distribution $\left( p_{i}^{\left( n\right) }\right) ,$ that is, 
$\varphi ^{\left( n\right) }=p^{\left( n\right) }=\sum_{i}p_{i}^{\left(
n\right) }\delta _{i},$ where for an initial value $x\in I$ and the
characteristic function $1_{A}$

\[
p_{i}^{\left( n\right) }\equiv \frac{1}{m+1}\sum_{k=n}^{m+n}1_{B_{i}}\left( 
\mathcal{F}^{k}x\right) . 
\]

\noindent When the initial value $x$ is distributed due to a measure $\nu $
on $I,$ the above $p_{i}^{\left( n\right) }$ is given as

\[
p_{i}^{\left( n\right) }\equiv \frac{1}{m+1}\int_{I}
\sum_{k=n}^{m+n}1_{B_{i}}\left( \mathcal{F}^{k}x\right) d\nu . 
\]

\noindent In the case that $\mathcal{F}$ is a classical baker's
transformation, if the orbit is not stable and periodic, then it is shown
that the $m\rightarrow \infty $ limit of $p_{i}^{\left( n\right) }$ exists
and equals to a natural invariant measure for a fixed $n\in \mathbf{N}$ \cite
{Ott}.

\noindent The joint distribution $\left( p_{ij}^{\left( n,n+1\right)
}\right) $ between the time $n$ and $n+1$ is defined by

\[
p_{ij}^{\left( n,n+1\right) }\equiv \frac{1}{m+1}\sum_{k=n}^{m+n}1_{B_{i}}
\left(\mathcal{F}^{k}x\right) 1_{B_{j}}\left(\mathcal{F}^{k+1}x\right) 
\]

\noindent or

\[
p_{ij}^{\left( n,n+1\right) }\equiv \frac{1}{m+1}\int_{I}
\sum_{k=n}^{m+n}1_{B_{i}}\left(\mathcal{F}^{k}x\right) 1_{B_{j}}\left( 
\mathcal{F}^{k+1}x\right) d\nu . 
\]

\noindent Then the channel $\Lambda _{n}^{*}$ at $n$ is determined by

\[
\Lambda _{n}^{*}\equiv \left( \frac{p_{ij}^{\left( n,n+1\right) }}{
p_{i}^{\left( n\right) }}\right) \Longrightarrow p^{\left( n+1\right)
}=\Lambda _{n}^{*}p^{\left( n\right) }, 
\]
and the chaos degree is given by

\begin{equation}
D_{c}\left( p^{\left( n\right) };\Lambda _{n}^{*}\right) =\sup_{\left\{
B_{k}\right\} }\left\{ \sum_{i}p_{i}^{\left( n\right) }S(\Lambda
_{n}^{*}\delta _{i})=\sum_{i,j}p_{ij}^{\left( n,n+1\right) }\log \frac{%
p_{i}^{\left( n\right) }}{p_{ij}^{\left( n,n+1\right) }};\left\{
B_{k}\right\} \right\} .  \label{4.2}
\end{equation}

\noindent We can judge whether the dynamics causes a chaos or not by the
value of D as

\begin{eqnarray*}
D &>&\mbox{0}\Longleftrightarrow \mbox{chaotic,} \\
D &=&\mbox{0}\Longleftrightarrow \mbox{stable.}
\end{eqnarray*}

\noindent Therefore it is enough to find a partition $\left\{ B_{k}\right\} $
such that $D$ is positive when the dynamics produces chaos.

This classical chaos degree was applied to several dynamical maps such
logistic map, Baker's transformation and Tinkerbel map, and it could explain
their chaotic characters\cite{O1,IOS}. Our chaos degree has several merits
compared with usual measures such as Lyapunov exponent.

\section{Expectation Values and Chaos Degree}

In this section, we show a general representation of the mean value of the
position operator $\hat{q}$ for the time evolution, which is constructed by
the quantum baker's map. Then we give the algorithm to compute the chaos
degree for the quantum baker's map.

To study the time evolution and the classical limit $h\rightarrow 0$ which
corresponds to $N\rightarrow \infty $ of the quantum baker's map $T$, we
introduce the following the mean value of the position operator $\hat{q}$
for time $n\in \mathbf{N}$ with respect to a single basis$\left| \xi
\right\rangle $:

\begin{equation}
r_{n}^{\left( N\right) }=\left\langle \xi \right| T^{n}\hat{q}T^{-n}\left|
\xi \right\rangle ,  \label{5.4}
\end{equation}

\noindent where $\left| \xi \right\rangle =\left| \xi _{1}\xi _{2}\cdots
\xi_{N}\right\rangle $ .

From (\ref{3.5}), the following formula of the matrix elements of $T^{n}$
for any $n\in \mathbf{N}$ is easily obtained.

\begin{eqnarray}
&&\left\langle \xi \right| T_{0}^{n}\left| \zeta \right\rangle  \nonumber \\
&=&\left\{ 
\begin{array}{ll}
\left( \frac{1-i}{2}\right)^{n}\left(\prod_{k=1}^{N-n}\delta \left( \xi
_{n+k}-\zeta _{k}\right) \right) \left( \prod_{l=1}^{n}A_{\xi _{l}\zeta
_{N-n+l}}\right) & \mbox{if } n<N \\ 
\left(\frac{1-i}{2}\right)^{n}\left(\prod_{k=1}^{n}A_{\xi _{k}\zeta
_{k}}\right) & \mbox{if}n=N \\ 
\left( \frac{1-i}{2}\right)^{n}\left( \prod\limits_{k=1}^{p}\left(
A^{m+1}\right) _{\xi _{k}\zeta _{N-p+k}}\right) \left(
\prod\limits_{l=1}^{N-p}\left( A^{m}\right) _{\xi _{p+l}\zeta _{l}}\right) & %
\mbox{if }n=mN+p \\ 
\left( \frac{1-i}{2}\right) ^{n}\prod\limits_{k=1}^{N}\left( A^{m}\right)
_{\xi _{k}\zeta _{k}} & \mbox{if }n=mN\mathbf{,}
\end{array}
\right.  \label{5.6}
\end{eqnarray}
\noindent where $A$ is the $2\times 2$ matrix with the element $%
A_{x_{1}x_{2}}=$ $\exp \left( \frac{\pi }{2}i\left| x_{1}-x_{2}\right|
\right) $ for $x_{1},x_{2}=0$ ,$1$, $p=1,\cdots ,N-1$ and $m\in \mathbf{N}$.

Using these formula, the following theorems are obtained and their proofs
are given in Appendix.

\begin{ftheo}
\begin{equation}
r_{n}^{\left( N\right) }=\left\{ 
\begin{array}{ll}
\sum_{k=1}^{N-n}\xi _{n+k}2^{-k}+\frac{2^{n}}{2^{N+1}} & \mbox{if }n<N \\ 
\frac{1}{2} & \mbox{if }n=N \\ 
\frac{1}{2^{n}}\sum_{j=0}^{2^{N}-1}\frac{j+1/2}{2^{N}}\prod\limits_{k=1}^{p}%
\left| \left( A^{m+1}\right) _{\xi _{k}j_{N-p+k}}\right|
^{2}\prod\limits_{l=1}^{N-p}\left| \left( A^{m}\right) _{\xi
_{p+l}j_{l}}\right| ^{2} & \mbox{if }n=mN+p \\ 
\frac{1}{2^{n}}\sum_{j=0}^{2^{N}-1}\frac{j+1/2}{2^{N}}\prod\limits_{k=1}^{N}%
\left| \left( A^{m}\right) _{\xi _{k}j_{k}}\right| ^{2} & \mbox{if }n=mN%
\mathbf{,}
\end{array}
\right.  \label{5.7}
\end{equation}
where $A$ is the $2\times 2$ matrix with the element $A_{x_{1}x_{2}}=$ $\exp
\left( \frac{\pi }{2}i\left| x_{1}-x_{2}\right| \right) $ for $x_{1},x_{2}=0$
,$1$, $p=1,\cdots ,N-1$ and $m\in \mathbf{N}$.
\end{ftheo}

\noindent By diagonalizing the matrix $A$, we obtain the following formula
of the absolute square of the matrix elements of $A^{n}$ for any $n\in 
\mathbf{N}$.

\begin{flemma}
For any $n\in \mathbf{N}$, we have 
\[
\left| \left( A^{n}\right) _{kj}\right| ^{2}=\left\{ 
\begin{array}{ll}
2^{n}\cos ^{2}\left( \frac{n\pi }{4}\right) & \mbox{if }k=j \\ 
2^{n}\sin ^{2}\left( \frac{n\pi }{4}\right) & \mbox{if }k\neq j
\end{array}
.\right. 
\]
\end{flemma}

\noindent \noindent Combining the above theorem and lemma, we obtain the
following two theorems with respect to the mean value $r_{n}^{\left(
N\right) }$ of the position operator.

\begin{ftheo}
For the case $n=mN+p,$ $p=1,2,\ldots N-1$ and $m\in \mathbf{N}$, we have

\begin{equation}
r_{n}^{\left( N\right) }=\left\{ 
\begin{array}{ll}
\sum_{k=1}^{N-p}\xi _{p+k}2^{-k}+\frac{2^{p}}{2^{N+1}} & \mbox{if }m=0\mbox{}%
\left( \mbox{mod}\mbox{ }4\right) \\ 
\sum_{k=N-p+1}^{N}\eta _{k-\left( N-p\right) }2^{-k}+\frac{2^{N}-2^{p}+1}{%
2^{N+1}} & \mbox{if }m=1\mbox{ }\left( \mbox{mod}\mbox{ }4\right) \\ 
\sum_{k=1}^{N-p}\eta _{p+k}2^{-k}+\frac{2^{p}}{2^{N+1}} & \mbox{if }m=2%
\mbox{ }\left( \mbox{mod}\mbox{ }4\right) \\ 
\sum_{k=N-p+1}^{N}\xi _{k-\left( N-p\right) }2^{-k}+\frac{2^{N}-2^{p}+1}{%
2^{N+1}} & \mbox{if }m=3\mbox{ }\left( \mbox{mod}\mbox{ }4\right) ,
\end{array}
\right.  \label{5.10}
\end{equation}
where $\eta _{k}=\xi _{k}+1\left( \mathrm{mod}\mbox{2}\right) ,k=1,\cdots ,N$
.
\end{ftheo}

\begin{ftheo}
For the case $n=mN,m\in \mathbf{N}$, we have 
\begin{equation}
r_{N}^{\left( n\right) }=\left\{ 
\begin{array}{ll}
\sum_{k=1}^{N}\xi _{k}2^{-k}+\frac{1}{2^{N+1}} & \mbox{if }m=0\left( \mathrm{%
mod}\mbox{ 4}\right) \\ 
\frac{1}{2} & \mbox{if }m=1,3\left( \mathrm{mod}\mbox{ 4}\right) \\ 
\sum_{k=1}^{N}\eta _{k}2^{-k}+\frac{1}{2^{N+1}} & \mbox{if }m=2\left( 
\mathrm{mod}\mbox{ 4}\right) .
\end{array}
\right.  \label{5.14}
\end{equation}
\end{ftheo}

Using these formulas (\ref{5.7}), (\ref{5.10}) and (\ref{5.14}) , the
probability distribution $\left( p_{i}^{\left( n\right) }\right) $ of the
orbit of mean value $r_{n}^{\left( N\right) }$ of the position operator $%
\hat{q}$ for the time evolution, which is constructed by the quantum baker's
map, is given by

\[
p_{i}^{\left( n\right) }\equiv \frac{1}{m+1}\sum_{k=n}^{m+n}1_{B_{i}}
\left(r_{n}^{\left( N\right) }\right) 
\]

\noindent for an initial value $r_{0}^{\left( N\right) }$ $\in \left[ 0,1%
\right] $ and the characteristic function $1_{A}.$ The joint distribution$%
\left( p_{ij}^{\left( n,n+1\right) }\right) $ between the time $n$ and $n+1$
is given by

\[
p_{ij}^{\left( n,n+1\right) }\equiv \frac{1}{m+1}\sum_{k=n}^{m+n}1_{B_{i}}
\left( r_{k}^{\left( N\right) }\right) 1_{B_{j}}\left( r_{k+1}^{\left(
N\right) }\right) . 
\]

\noindent Thus the chaos degree for the quantum baker's map is calculated by

\begin{equation}
D_{q}\left( p^{\left( n\right) };\Lambda
_{n}^{*}\right)=\sum_{i,j}p_{ij}^{\left( n,n+1\right) }\log \frac{%
p_{i}^{\left( n\right) }}{p_{ij}^{\left( n,n+1\right) }},
\end{equation}

\noindent whose numerical value is shown in the next section.

\section{Numerical Simulation of the Chaos Degree and Classical-Quantum
Correspondence}

We compare the dynamics of the mean value $r_{n}^{\left( N\right) }$ of
position operator $\hat{q}$ with that of the classical value $q_{n}$ in the $%
q$ direction. We take an initial value of the mean value as

\[
r_{0}^{\left( N\right) }=\sum_{l=1}^{N}\xi _{l}2^{-l}+1/2^{N+1}=0.\xi_{1}\xi
_{2}\cdots \xi _{N}1, 
\]

\noindent where $\xi _{i}$ is a pseudo-random number valued with $0$ or $1$.
At the time zero we assume that the classical value $q_{0}$ in the $q$
direction takes the same value as the mean value $r_{0}^{\left( N\right) }$
of position operator $\hat{q}$. The distribution of $r_{n}^{\left( N\right)} 
$ for the case $N=500$ is shown in Fig.1 up to the time $n=1000$. The
distribution of the classical value $q_{n}$ for the case $N=500$ in the $q$
direction is shown in Fig.2 up to the time $n=1000$.

\begin{center}
\includegraphics[width=10.0cm,height=7.0cm] {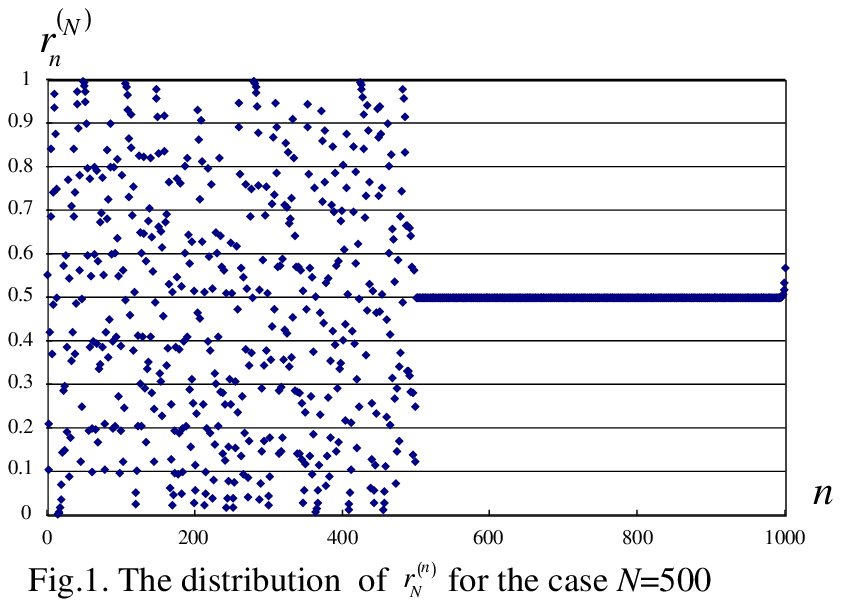} %
\includegraphics[width=10.0cm,height=7.0cm] {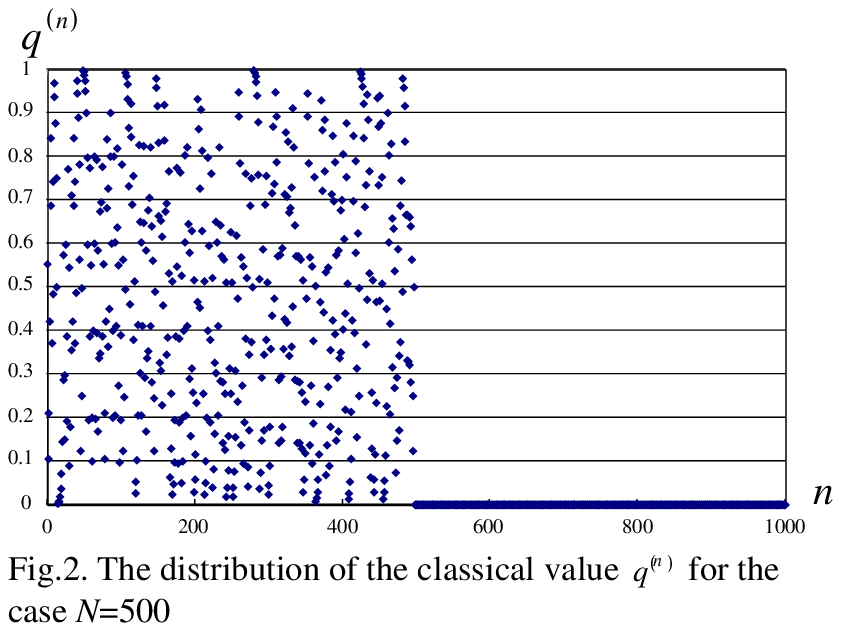}
\end{center}

Fig.3 presents the change of the chaos degree for the case $%
N=100,300,500,700 $ up to the time $n=1000$.

\begin{center}
\includegraphics[width=10.0cm,height=7.0cm] {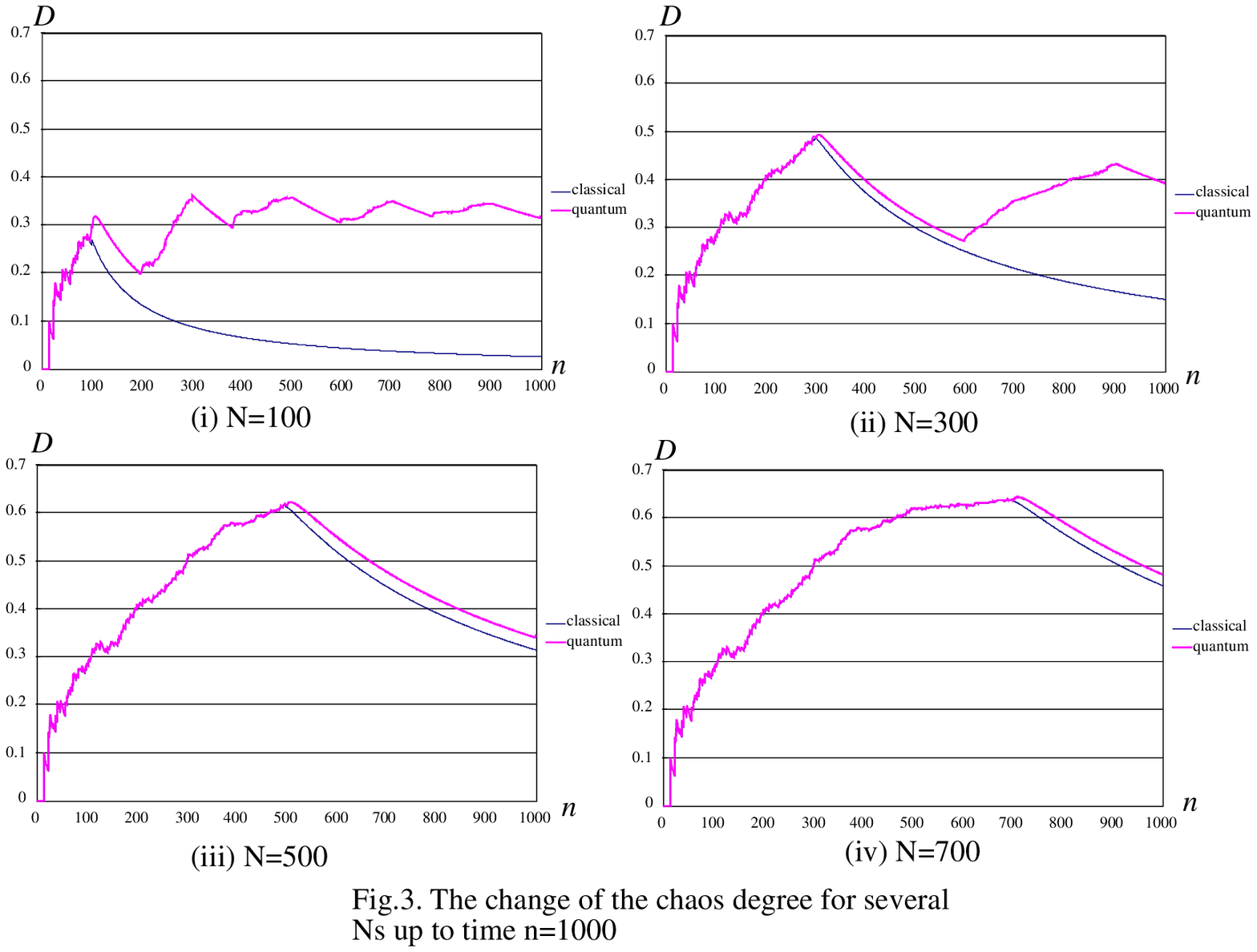}
\end{center}

The correspondence between the chaos degree $D_{q}$ for the quantum baker's
map and the chaos degree $D_{c}$ for the classical baker's map for some
fixed $Ns$ ($100,300,500,700$ here) is shown for the time less than $T=\log
_{2}\frac{1}{h}=\log _{2}2^{N}=N$, and it is lost at the logarithtic time
scale $T$. Here we took a finite partition $\left\{ B_{k}\right\} $ of $I=%
\left[ 0,1\right] $ such as $B_{k}=\left[ \frac{k}{100},\frac{k+1}{100}
\right) \left( k=0,1,\ldots ,98\right) $ and $B_{99}=\left[ \frac{99}{100} ,1%
\right] $ to compute the chaos degree numerically.

The difference of the chaos degrees between the chaos degree $D_{q}$ for the
quantum baker's map and the chaos degree $D_{c}$ for the classical baker's
map for a fixed time $n$ ($1000$, here) is displayed w.r.t. $N$ in Fig.4.

\begin{center}
\includegraphics[width=10.0cm,height=7.0cm] {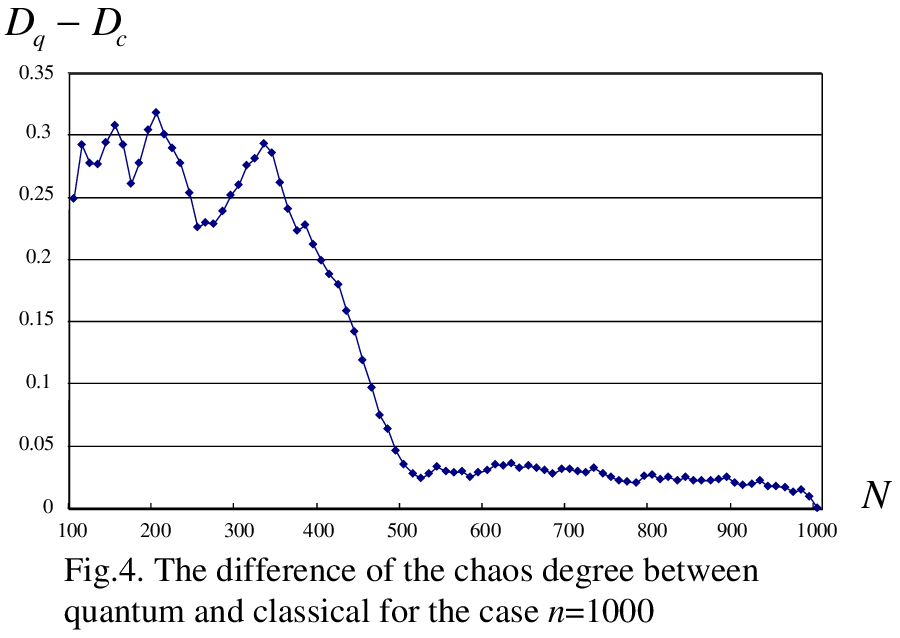}
\end{center}

\noindent Thus we conclude that the dynamics of the mean value $q_{n}$
reduces the classical dynamics $q_{n}$ in the $q$ direction in the classical
limit $N\rightarrow \infty \left( h\rightarrow 0\right) $.

The appearance of the logarithmic timescale have been proved rigorously in
our recent paper \cite{IOV}.

\section{Acknowledgments}

The main part of this work was done during the visit of I.V. to the Science
University of Tokyo. He (I.V.) is grateful to JSPS for the Fellowship award.
Our work was also partially supported by RFFI 99-0100866, INTAS 99-00545 and
SCAT.

\section{Appendix}

\noindent \textbf{Proof of THEOREM 5.1:} By a direct calculation, we obtain

\begin{eqnarray*}
r_{n}^{\left( N\right) } &=&\left\langle \xi \right| T^{n}\hat{q}%
T^{-n}\left| \xi \right\rangle \\
&=&\left\langle \xi \right| T^{n}\left( \sum_{j=0}^{2^{N}-1}\frac{j+1/2}{%
2^{N}}\left| j\right\rangle \left\langle j\right| \right) T^{-n}\left| \xi
\right\rangle \\
&=&\sum_{j=0}^{2^{N}-1}\frac{j+1/2}{2^{N}}\left\langle \xi \right|
T^{n}\left| j\right\rangle \left\langle j\right| T^{-n}\left| \xi
\right\rangle \\
&=&\sum_{j=0}^{2^{N}-1}\frac{j+1/2}{2^{N}}\left\langle \xi \right|
T^{n}\left| j\right\rangle \left\langle j\right| T^{*n}\left| \xi
\right\rangle \\
&=&\sum_{j=0}^{2^{N}-1}\frac{j+1/2}{2^{N}}\left\langle \xi \right|
T^{n}\left| j\right\rangle \overline{\left\langle \xi \right| T^{n}\left|
j\right\rangle } \\
&=&\sum_{j=0}^{2^{N}-1}\frac{j+1/2}{2^{N}}\left| \left\langle \xi \right|
T^{n}\left| j\right\rangle \right|^{2}.
\end{eqnarray*}

\noindent Using (\ref{5.6}) the mean value $r_{n}^{\left( N\right) }$ in the
case $n<N$ can be expressed as

\begin{eqnarray*}
r_{n}^{\left( N\right) } &=&\sum_{j=0}^{2^{N}-1}\frac{j+1/2}{2^{N}}\left|
\left\langle \xi \right| T^{n}\left| j\right\rangle \right| ^{2}. \\
&=&\sum_{j=0}^{2^{N}-1}\frac{j+1/2}{2^{N}}\left| \left( \frac{1-i}{2}%
\right)^{n}\left( \prod_{k=1}^{N-n}\delta \left( \xi _{n+k}-j_{k}\right)
\right) \left( \prod_{l=1}^{n}A_{\xi _{l}j_{N-n+l}}\right) \right| ^{2} \\
&=&\sum_{j=0}^{2^{N}-1}\frac{j+1/2}{2^{N}}\left( \frac{1-i}{2}%
\right)^{n}\left( \prod_{k=1}^{N-n}\delta \left( \xi _{n+k}-j_{k}\right)
\right) \left( \prod_{l=1}^{n}A_{\xi _{l}j_{N-n+l}}\right) \\
& &\times \overline{\left( \frac{1-i}{2}\right)
^{n}\left(\prod_{k=1}^{N-n}\delta \left( \xi _{n+k}-j_{k}\right) \right)
\left( A_{\xi_{l}j_{N-n+l}}\right) } \\
&=&\sum_{j=0}^{2^{N}-1}\frac{j+1/2}{2^{N}}\left( \frac{1-i}{2}%
\right)^{n}\left( \frac{1+i}{2}\right) ^{n}\left( \prod_{k=1}^{N-n}\delta
\left( \xi _{n+k}-j_{k}\right) \right) \left( \prod_{l=1}^{n}\left|
A_{\xi_{l}j_{N-n+l}}\right| ^{2}\right) \\
&=&\sum_{j=0}^{2^{N}-1}\frac{j+1/2}{2^{N}}\left( \frac{1-i}{2}%
\right)^{n}\left( \frac{1+i}{2}\right) ^{n}\left( \prod_{k=1}^{N-n}\delta
\left( \xi _{n+k}-j_{k}\right) \right) \\
&=&\frac{1}{2^{N+n}}\sum_{j_{1},\cdots j_{N}}\left\{ \left(
\sum_{k=1}^{N}j_{k}2^{N-k}\right) +1/2\right\} \left(
\prod_{k=1}^{N-n}\delta \left( \xi _{n+k}-j_{k}\right) \right) \\
&=&\frac{1}{2^{N+n}}\sum_{j_{1},\cdots j_{N}}\left(
\sum_{k=1}^{N}j_{k}2^{N-k}\right) \left( \prod_{k=1}^{N-n}\delta \left(
\xi_{n+k}-j_{k}\right) \right) \\
&&+\frac{1}{2^{N+n+1}}\sum_{j_{1},\cdots j_{N}}\left(
\prod_{k=1}^{N-n}\delta \left( \xi _{n+k}-j_{k}\right) \right) \\
&=&\frac{1}{2^{N+n}}\sum_{j_{N-n+1},\cdots j_{N}}\left(
\sum_{l=1}^{N-n}\xi_{n+l}2^{N-l}+\sum_{l=N-n+1}^{N}j_{l}2^{N-l}\right) +%
\frac{1}{2^{N+n+1}}\left( \sum_{j_{N-n+1},\cdots j_{N}}1\right) \\
&=&\frac{1}{2^{N+n}}\left( \sum_{l=1}^{N-n}\xi _{n+l}2^{N-l}\right) \left(
\sum_{j_{N-n+1},\cdots j_{N}}1\right) +\frac{1}{2^{N+n}}\sum_{j_{N-n+1},%
\cdots j_{N}}\left( \sum_{l=N-n+1}^{N}j_{l}2^{N-l}\right) \\
&&+\frac{1}{2^{N+n+1}}\left( \sum_{j_{N-n+1},\cdots j_{N}}1\right) \\
&=&\frac{2^{n}}{2^{N+n}}\left( \sum_{l=1}^{N-n}\xi _{n+l}2^{N-l}\right) +%
\frac{1}{2^{N+n}}\sum_{j_{N-n+1},\cdots j_{N}}\left(
\sum_{l=N-n+1}^{N}j_{l}2^{N-l}\right) +\frac{2^{n}}{2^{N+n+1}} \\
&=&\frac{1}{2^{N}}\left( \sum_{l=1}^{N-n}\xi _{n+l}2^{N-l}\right) +\frac{1}{%
2^{N+n}}\sum_{j_{N-n+1},\cdots j_{N}}\left(
\sum_{l=1}^{n}j_{N-n+l}2^{n-l}\right) +\frac{1}{2^{N+1}} \\
&=&\frac{1}{2^{N}}\left( \sum_{l=1}^{N-n}\xi _{n+l}2^{N-l}\right) +\frac{1}{%
2^{N+n}}\frac{1}{2}\left( 2^{n}-1\right) 2^{n}+\frac{1}{2^{N+1}} \\
&=&\frac{1}{2^{N}}\left( \sum_{l=1}^{N-n}\xi _{n+l}2^{N-l}\right) +\frac{
2^{n}}{2^{N+1}}
\end{eqnarray*}

\noindent For the case $n=N$, we similarly obtain

\begin{eqnarray*}
r_{n}^{\left( N\right) } &=&\sum_{j=0}^{2^{N}-1}\frac{j+1/2}{2^{N}}\left|
\left\langle \xi \right| T^{n}\left| j\right\rangle \right| ^{2}. \\
&=&\sum_{j=0}^{2^{N}-1}\frac{j+1/2}{2^{N}}\left| \left( \frac{1-i}{2}
\right)^{N}\left( \prod_{k=1}^{N}A_{\xi _{k}j_{k}}\right) \right| ^{2} \\
&=&\sum_{j=0}^{2^{N}-1}\frac{j+1/2}{2^{N}}\left| \left( \frac{1-i}{2}
\right)^{N}\right| ^{2}\prod_{k=1}^{N}\left| A_{\xi _{k}j_{k}}\right| ^{2} \\
&=&\frac{1}{2^{2N}}\sum_{j=0}^{2^{N}-1}\left( j+1/2\right) \\
&=&\frac{1}{2^{2N}}\frac{1}{2}\left( 2^{N}-1\right) 2^{N}+\frac{1}{2^{N+1}}
\\
&=&\frac{1}{2}.
\end{eqnarray*}

For $n=mN+p,p=1,2,\cdots ,N-1\ $, $m\in \mathbf{N}$,

\begin{eqnarray*}
r_{n}^{\left( N\right) } &=&\sum_{j=0}^{2^{N}-1}\frac{j+1/2}{2^{N}}\left|
\left\langle \xi \right| T^{n}\left| j\right\rangle \right| ^{2} \\
&=&\sum_{j=0}^{2^{N}-1}\frac{j+1/2}{2^{N}}\left| \left( \frac{1-i}{2}%
\right)^{n}\left( \prod\limits_{k=1}^{p}\left( A^{m+1}\right)
_{\xi_{k}j_{N-p+k}}\right) \left( \prod\limits_{l=1}^{N-p}\left(
A^{m}\right)_{\xi _{p+l}j_{l}}\right) \right| ^{2} \\
&=&\frac{1}{2^{n}}\sum_{j=0}^{2^{N}-1}\frac{j+1/2}{2^{N}}\prod%
\limits_{k=1}^{p}\left| \left( A^{m+1}\right) _{\xi _{k}j_{N-p+k}}\right|
^{2}\prod\limits_{l=1}^{N-p}\left| \left( A^{m}\right) _{\xi
_{p+l}j_{l}}\right| ^{2}
\end{eqnarray*}

\noindent and for $n=mN,$ $m\in \mathbf{N}$,

\begin{eqnarray*}
r_{N}^{\left( n\right) } &=&\sum_{j=0}^{2^{N}-1}\frac{j+1/2}{2^{N}}\left|
\left\langle \xi \right| T^{n}\left| j\right\rangle \right| ^{2} \\
&=&\sum_{j=0}^{2^{N}-1}\frac{j+1/2}{2^{N}}\left| \left( \frac{1-i}{2}%
\right)^{n}\prod\limits_{k=1}^{N}\left( A^{m}\right) _{\xi _{k}j_{k}}\right|
^{2} \\
&=&\frac{1}{2^{n}}\sum_{j=0}^{2^{N}-1}\frac{j+1/2}{2^{N}}\prod%
\limits_{k=1}^{N}\left| \left( A^{m}\right) _{\xi _{k}j_{k}}\right| ^{2}.
\end{eqnarray*}

$\hspace{15cm}\blacksquare $

\noindent \textbf{Proof of LEMMA 5.2: }By a direct calculation, the matrix $%
A $ is diagonalized as follows: 
\begin{equation}
A=FDF^{*},  \label{5.8}
\end{equation}
\noindent where 
\[
F=\frac{1}{\sqrt{2}}\left( 
\begin{array}{ll}
1 & -1 \\ 
1 & 1
\end{array}
\right) ,D=\left( 
\begin{array}{ll}
1+i & 0 \\ 
0 & 1-i
\end{array}
\right) . 
\]
\noindent From (\ref{5.8}), we have

\begin{eqnarray}
A^{n} &=&FD^{n}F^{*}  \nonumber \\
&=&\frac{1}{\sqrt{2}}\left( 
\begin{array}{ll}
1 & -1 \\ 
1 & 1
\end{array}
\right) \left( 
\begin{array}{ll}
\left( 1+i\right) ^{n} & 0 \\ 
0 & \left( 1-i\right) ^{n}
\end{array}
\right) \frac{1}{\sqrt{2}}\left( 
\begin{array}{ll}
1 & 1 \\ 
-1 & 1
\end{array}
\right)  \nonumber \\
&=&\frac{1}{2}\left( 
\begin{array}{ll}
\left( 1+i\right) ^{n}+\left( 1-i\right) ^{n} & \left( 1+i\right)^{n}-\left(
1-i\right) ^{n} \\ 
\left( 1+i\right) ^{n}-\left( 1-i\right) ^{n} & \left( 1+i\right)^{n}+\left(
1-i\right) ^{n}
\end{array}
\right) .  \label{fn}
\end{eqnarray}

\noindent Using (\ref{fn}), it follows that for any $k=j,k=1,2$,

\begin{eqnarray*}
\left| \left( A^{n}\right) _{kj}\right| ^{2} &=&\frac{1}{2}\left\{ \left(
1+i\right) ^{n}+\left( 1-i\right) ^{n}\right\} \overline{\frac{1}{2}\left\{
\left( 1+i\right) ^{n}+\left( 1-i\right) ^{n}\right\} } \\
&=&\frac{1}{4}\left\{ \left( 1+i\right) ^{n}+\left( 1-i\right) ^{n}\right\}
\left\{ \left( 1-i\right) ^{n}+\left( 1+i\right) ^{n}\right\} \\
&=&\frac{1}{4}\left\{ \left( 1+i\right) ^{n}+\left( 1-i\right) ^{n}\right\}
^{2} \\
&=&\frac{1}{4}\left\{ \left( \sqrt{2}\frac{1+i}{\sqrt{2}}\right) ^{n}+\left(%
\sqrt{2}\frac{1-i}{\sqrt{2}}\right) ^{n}\right\} ^{2} \\
&=&\frac{1}{4}\left\{ \left( \sqrt{2}\right) ^{n}\left( \frac{1+i}{\sqrt{2}}%
\right) ^{n}+\left( \sqrt{2}\right) ^{n}\left( \frac{1-i}{\sqrt{2}}%
\right)^{n}\right\} ^{2} \\
&=&\frac{2^{n}}{4}\left\{ \left( \exp \left( \frac{\pi }{4}i\right)
\right)^{n}+\left( \exp \left( -\frac{\pi }{4}i\right) \right) ^{n}\right\}
^{2} \\
&=&\frac{2^{n}}{4}\left\{ \exp \left( \frac{n\pi }{4}i\right) +\exp \left( -%
\frac{n\pi }{4}i\right) \right\} ^{2} \\
&=&\frac{2^{n}}{4}\left[ \left\{ \cos \left( \frac{n\pi }{4}\right) +i\sin
\left( \frac{n\pi }{4}\right) \right\} +\left\{ \cos \left( \frac{n\pi }{4}%
\right) -i\sin \left( \frac{n\pi }{4}\right) \right\} \right] ^{2} \\
&=&\frac{2^{n}}{4}\left\{ 2\cos \left( \frac{n\pi }{4}\right) \right\} ^{2}
\\
&=&2^{n}\cos ^{2}\left( \frac{n\pi }{4}\right)
\end{eqnarray*}

\noindent and for any $k\neq j,k=1,2$,

\begin{eqnarray*}
\left| \left( A^{n}\right) _{kj}\right| ^{2} &=&\frac{1}{2}\left\{ \left(
1+i\right) ^{n}-\left( 1-i\right) ^{n}\right\} \overline{\frac{1}{2}\left\{
\left( 1+i\right) ^{n}-\left( 1-i\right) ^{n}\right\} } \\
&=&\frac{1}{4}\left\{ \left( 1+i\right) ^{n}-\left( 1-i\right) ^{n}\right\}
\left\{ \left( 1-i\right) ^{n}-\left( 1+i\right) ^{n}\right\} \\
&=&-\frac{1}{4}\left\{ \left( 1+i\right) ^{n}-\left( 1-i\right)
^{n}\right\}^{2} \\
&=&-\frac{1}{4}\left\{ \left( \sqrt{2}\frac{1+i}{\sqrt{2}}\right)^{n}-\left( 
\sqrt{2}\frac{1-i}{\sqrt{2}}\right) ^{n}\right\} ^{2} \\
&=&-\frac{1}{4}\left\{ \left( \sqrt{2}\right) ^{n}\left( \frac{1+i}{\sqrt{2}}%
\right) ^{n}-\left( \sqrt{2}\right) ^{n}\left( \frac{1-i}{\sqrt{2}}%
\right)^{n}\right\} ^{2} \\
&=&-\frac{2^{n}}{4}\left\{ \left( \exp \left( \frac{\pi }{4}i\right)
\right)^{n}-\left( \exp \left( -\frac{\pi }{4}i\right) \right) ^{n}\right\}
^{2} \\
&=&-\frac{2^{n}}{4}\left[ \left\{ \cos \left( \frac{n\pi }{4}\right) +i\sin
\left( \frac{n\pi }{4}\right) \right\} -\left\{ \cos \left( \frac{n\pi }{4}%
\right) -i\sin \left( \frac{n\pi }{4}\right) \right\} \right] ^{2} \\
&=&-\frac{2^{n}}{4}\left\{ 2i\sin \left( \frac{n\pi }{4}\right) \right\} ^{2}
\\
&=&2^{n}\sin ^{2}\left( \frac{n\pi }{4}\right)
\end{eqnarray*}

$\hspace{15cm}\blacksquare $ \noindent \noindent

\noindent \textbf{Proof of THEOREM 5.3:} For the case $n=mN+p,$ $p=1,\cdots
,N-1$ and $m\in \mathbf{N}$,

\[
r_{n}^{\left( N\right) }=\frac{1}{2^{n}}\sum_{j=0}^{2^{N}-1}\frac{j+1/2}{%
2^{N}}\prod\limits_{k=1}^{p}\left| \left( A^{m+1}\right) _{\xi
_{k}j_{N-p+k}}\right| ^{2}\prod\limits_{l=1}^{N-p}\left| \left(
A^{m}\right)_{\xi _{p+l}j_{l}}\right| ^{2}. 
\]

\noindent By a direct calculation, we obtain

\begin{eqnarray}
r_{n}^{\left( N\right) } &=&\frac{1}{2^{n}}\sum_{j=0}^{2^{N}-1}\frac{j+1/2}{%
2^{N}}\prod\limits_{k=1}^{p}\left| \left( A^{m+1}\right)
_{\xi_{k}j_{N-p+k}}\right| ^{2}\prod\limits_{l=1}^{N-p}\left| \left(
A^{m}\right)_{\xi _{p+l}j_{l}}\right| ^{2}  \nonumber \\
&=&\frac{1}{2^{n}}\sum_{j=0}^{2^{N}-1}\frac{j+1/2}{2^{N}}\prod%
\limits_{l=1}^{N-p}\left| \left( A^{m}\right) _{\xi _{p+l}j_{l}}\right|
^{2}\prod\limits_{k=1}^{p}\left| \left( A^{m+1}\right)
_{\xi_{k}j_{N-p+k}}\right| ^{2}  \nonumber \\
&=&\frac{1}{2^{n+N}}\sum_{j=0}^{2^{N}-1}\left( j+\frac{1}{2}\right)
\prod\limits_{l=1}^{N-p}\left| \left( A^{m}\right) _{\xi
_{p+l}j_{l}}\right|^{2}\prod\limits_{k=1}^{p}\left| \left( A^{m+1}\right)
_{\xi _{k}j_{N-p+k}}\right| ^{2}  \nonumber \\
&=&\frac{1}{2^{n+N}}\sum_{j_{1},\cdots ,j_{N}}\left\{ \left(
\sum_{k=1}^{N}j_{k}2^{N-k}\right) +\frac{1}{2}\right\}
\prod\limits_{l=1}^{N-p}\left| \left( A^{m}\right) _{\xi _{p+l}j_{l}}\right|
^{2}\prod\limits_{k=1}^{p}\left| \left( A^{m+1}\right) _{\xi
_{k}j_{N-p+k}}\right| ^{2}  \nonumber \\
&=&\frac{1}{2^{n+N}}\sum_{j_{1},\cdots ,j_{N}}\left\{ \left(
\sum_{k=1}^{N-p}j_{k}2^{N-k}\right) +\left(
\sum_{k=N-p+1}^{N}j_{k}2^{N-k}\right) +\frac{1}{2}\right\}  \nonumber \\
&&\times \prod\limits_{l=1}^{N-p}\left| \left( A^{m}\right) _{\xi
_{p+l}j_{l}}\right| ^{2}\prod\limits_{k=1}^{p}\left| \left( A^{m+1}\right)
_{\xi _{k}j_{N-p+k}}\right| ^{2}  \nonumber \\
&=&\frac{1}{2^{\left( m+1\right) N+p}}\sum_{j_{1},\cdots ,j_{N}}\left\{
\left( \sum_{k=1}^{N-p}j_{k}2^{N-k}\right) +\left(
\sum_{k=N-p+1}^{N}j_{k}2^{N-k}\right) +\frac{1}{2}\right\}  \nonumber \\
&&\times \prod\limits_{l=1}^{N-p}\left| \left( A^{m}\right)
_{\xi_{p+l}j_{l}}\right| ^{2}\prod\limits_{k=N-p+1}^{N}\left|
\left(A^{m+1}\right) _{\xi _{k-\left( N-p\right) }j_{k}}\right| ^{2}
\label{5.11}
\end{eqnarray}

\noindent (i) $m=0\left( \mathrm{mod} \mbox{4}\right) $

\noindent From the above lemma, we have 
\[
\left| \left( A^{m}\right) _{\xi _{p+l}j_{l}}\right| ^{2}=\left\{ 
\begin{array}{ll}
2^{m} & \mbox{if }j_{l}=\xi _{p+l} \\ 
0 & \mbox{if }j_{l}\neq j_{p+l}
\end{array}
\right. ,\;\left| \left( A^{m+1}\right) _{\xi _{k-\left( N-p\right)
j_{k}}}\right| ^{2}=2^{m} 
\]

\noindent for any $l=1,\cdots ,N-p$ and $k=N-p+1,\cdots ,N$. Using this
formula the product of absolute squares can be expressed as

\begin{eqnarray*}
&&\prod\limits_{l=1}^{N-p}\left| \left( A^{m}\right)
_{\xi_{p+l}j_{l}}\right| ^{2}\prod\limits_{k=N-p+1}^{N}\left|
\left(A^{m+1}\right) _{\xi _{k-\left( N-p\right)}j_{k}}\right| \\
&=&\left\{ 
\begin{array}{ll}
\left( 2^{m}\right) ^{N-p}\left( 2^{m}\right) ^{p} & \mbox{if }%
j_{l}=\xi_{p+l}\mbox{ for all }l=1,\cdots ,N-p \\ 
0 & \mbox{otherwise}
\end{array}
\right. \\
&=&\left\{ 
\begin{array}{ll}
2^{mN} & \mbox{if }j_{l}=\xi _{p+l}\mbox{ for all }l=1,\cdots ,N-p \\ 
0 & \mbox{otherwise}
\end{array}
\right. .
\end{eqnarray*}

\noindent (\ref{5.11}) can be rewritten as

\begin{eqnarray}
r_{n}^{\left( N\right) } &=&\frac{1}{2^{\left( m+1\right) N+p}}
\sum_{j_{1},\cdots ,j_{N}}\left\{ \left(
\sum_{k=1}^{N-p}j_{k}2^{N-k}\right)+\left(
\sum_{k=N-p+1}^{N}j_{k}2^{N-k}\right) +\frac{1}{2}\right\}  \nonumber \\
&&\times \prod\limits_{l=1}^{N-p}\left| \left( A^{m}\right)
_{\xi_{p+l}j_{l}}\right| ^{2}\prod\limits_{k=N-p+1}^{N}\left|
\left(A^{m+1}\right) _{\xi _{k-\left( N-p\right) }j_{k}}\right| ^{2} 
\nonumber \\
&=&\frac{2^{mN}}{2^{\left( m+1\right) N+p}}\sum_{j_{N-p+1},\cdots
,j_{N}}\left\{ \left( \sum_{k=1}^{N-p}\xi _{p+k}2^{N-k}\right) +\left(
\sum_{k=N-p+1}^{N}j_{k}2^{N-k}\right) +\frac{1}{2}\right\}  \label{5.12} \\
&=&\frac{1}{2^{N+p}}\left( \sum_{k=1}^{N-p}\xi _{p+k}2^{N-k}\right)
\left(\sum_{j_{N-p+1},\cdots ,j_{N}}1\right) +\frac{1}{2^{N+p}}
\sum_{j_{N-p+1},\cdots ,j_{N}}\left( \sum_{k=N-p+1}^{N}j_{k}2^{N-k}\right) 
\nonumber \\
&&+\frac{1}{2^{N+p}}\frac{1}{2}\left( \sum_{j_{N-p+1},\cdots ,j_{N}}1\right)
\nonumber \\
&=&\frac{1}{2^{N}}\sum_{k=1}^{N-p}\xi _{p+k}2^{N-k}+\frac{1}{2^{N+p}}%
\sum_{j_{N-p+1},\cdots ,j_{N}}\left( \sum_{k=N-p+1}^{N}j_{k}2^{N-k}\right) +%
\frac{1}{2^{N+1}}  \nonumber \\
&=&\frac{1}{2^{N}}\sum_{k=1}^{N-p}\xi _{p+k}2^{N-k}+\frac{1}{2^{N+p}}
\sum_{j_{N-p+1},\cdots ,j_{N}}\left( \sum_{k=1}^{p}j_{N-p+k}2^{p-k}\right) +%
\frac{1}{2^{N+1}}  \nonumber \\
&=&\frac{1}{2^{N}}\sum_{k=1}^{N-p}\xi _{p+k}2^{N-k}+\frac{1}{2^{N+p}}%
\sum_{k=0}^{2^{p}-1}k+\frac{1}{2^{N+1}}  \nonumber \\
&=&\frac{1}{2^{N}}\sum_{k=1}^{N-p}\xi _{p+k}2^{N-k}+\frac{1}{2^{N+p}}\frac{1%
}{2}\left( 2^{p}-1\right) 2^{p}+\frac{1}{2^{N+1}}  \nonumber \\
&=&\sum_{k=1}^{N-p}\xi _{p+k}2^{-k}+\frac{2^{p}}{2^{N+1}}.  \nonumber
\end{eqnarray}

\noindent (ii) $m=1\left( \mathrm{mod} 4\right) $

\noindent From the above lemma, we have

\[
\left| \left( A^{m}\right) _{\xi _{p+l}j_{l}}\right| ^{2}=2^{m-1},\; \left|
\left( A^{m+1}\right) _{\xi _{k-\left( N-p\right) j_{k}}}\right|
^{2}=\left\{ 
\begin{array}{ll}
2^{m+1} & \mbox{if }j_{k}\neq \xi _{k-\left( N-p\right) } \\ 
0 & \mbox{if }j_{k}=\xi _{k-\left( N-p\right) }
\end{array}
\right. 
\]

\noindent for any $l=1,\cdots ,N-p$ and $k=N-p+1,\cdots ,N$. Using this
formula the product of absolute squares can be expressed as

\begin{eqnarray*}
&&\prod\limits_{l=1}^{N-p}\left| \left( A^{m}\right)
_{\xi_{p+l}j_{l}}\right| ^{2}\prod\limits_{k=N-p+1}^{N}\left|
\left(A^{m+1}\right) _{\xi _{k-\left( N-p\right) }j_{k}}\right| \\
&=&\left\{ 
\begin{array}{ll}
\left( 2^{m-1}\right) ^{N-p}\left( 2^{m+1}\right) ^{p} & \mbox{if }%
j_{k}\neq\xi _{k-\left( N-p\right) }\mbox{ for all }k=N-p+1,\cdots ,N \\ 
0 & \mbox{otherwise}
\end{array}
\right. \\
&=&\left\{ 
\begin{array}{ll}
2^{\left( m-1\right) N+2p} & \mbox{if }j_{k}\neq \xi _{k-\left( N-p\right) }%
\mbox{ for all }k=N-p+1,\cdots ,N \\ 
0 & \mbox{otherwise}
\end{array}
\right. .
\end{eqnarray*}

\noindent Let $\eta _{k-\left( N-p\right) }=\xi _{k-\left(
N-p\right)}+1\left( \mathrm{mod} \mbox{ 2}\right) ,k=N-p+1,\cdots ,N$. It
follows that

\begin{eqnarray}
r_{n}^{\left( N\right) } &=&\frac{1}{2^{\left( m+1\right) N+p}}
\sum_{j_{1},\cdots ,j_{N}}\left\{ \left(
\sum_{k=1}^{N-p}j_{k}2^{N-k}\right)+\left(
\sum_{k=N-p+1}^{N}j_{k}2^{N-k}\right) +\frac{1}{2}\right\}  \nonumber \\
&&\times \prod\limits_{l=1}^{N-p}\left| \left( A^{m}\right)
_{\xi_{p+l}j_{l}}\right| ^{2}\prod\limits_{k=N-p+1}^{N}\left|
\left(A^{m+1}\right) _{\xi _{k-\left( N-p\right) }j_{k}}\right| ^{2} 
\nonumber \\
&=&\frac{2^{\left( m-1\right) N+2p}}{2^{\left( m+1\right) N+p}}%
\sum_{j_{1},\cdots ,j_{N-p}}\left\{
\left(\sum_{k=1}^{N-p}j_{k}2^{N-k}\right) +\left(
\sum_{k=N-p+1}^{N}\eta_{k-\left( N-p\right) }2^{N-k}\right) +\frac{1}{2}%
\right\}  \label{5.13} \\
&=&\frac{1}{2^{2N-p}}\sum_{j_{1},\cdots ,j_{N-p}}\left\{ \left(
\sum_{k=1}^{N-p}j_{k}2^{N-k}\right) +\left( \sum_{k=N-p+1}^{N}\eta
_{k-\left( N-p\right) }2^{N-k}\right) +\frac{1}{2}\right\}  \nonumber \\
&=&\frac{1}{2^{2N-p}}\sum_{j_{1},\cdots
,j_{N-p}}\left(\sum_{k=1}^{N-p}j_{k}2^{N-k}\right) +\frac{1}{2^{2N-p}}%
\left(\sum_{k=N-p+1}^{N}\eta _{k-\left( N-p\right) }2^{N-k}\right)
\left(\sum_{j_{1},\cdots ,j_{N-p}}1\right)  \nonumber \\
&&+\frac{1}{2^{2N-p}}\frac{1}{2}\left( \sum_{j_{1},\cdots ,j_{N-p}}1\right) 
\nonumber \\
&=&\frac{1}{2^{2N-p}}\sum_{j_{1},\cdots ,j_{N-p}}\left(
\sum_{k=1}^{N-p}j_{k}2^{N-k}\right) +\frac{2^{N-p}}{2^{2N-p}}\left(
\sum_{k=N-p+1}^{N}\eta _{k-\left( N-p\right) }2^{N-k}\right) +\frac{2^{N-p}}{%
2^{2N-p}}\frac{1}{2}  \nonumber \\
&=&\frac{1}{2^{2N-p}}\sum_{j_{1},\cdots ,j_{N-p}}\left(
\sum_{k=1}^{N-p}j_{k}2^{N-k}\right) +\sum_{k=N-p+1}^{N}\eta _{k-\left(
N-p\right) }2^{-k}+\frac{1}{2^{N+1}}  \nonumber \\
&=&\frac{2^{p}}{2^{2N-p}}\sum_{j_{1},\cdots ,j_{N-p}}\left(
\sum_{k=1}^{N-p}j_{k}2^{N-p-k}\right) +\sum_{k=N-p+1}^{N}\eta
_{k-\left(N-p\right) }2^{-k}+\frac{1}{2^{N+1}}  \nonumber \\
&=&\frac{2^{p}}{2^{2N-p}}\sum_{k=0}^{2^{N-p}-1}k+\sum_{k=N-p+1}{N}
\eta_{k-\left( N-p\right) }2^{-k}+\frac{1}{2^{N+1}}  \nonumber \\
&=&\frac{2^{p}}{2^{2N-p}}\frac{1}{2}\left( 2^{N-p}-1\right)
2^{N-p}+\sum_{k=N-p+1}^{N}\eta _{k-\left( N-p\right) }2^{-k}+\frac{1}{2^{N+1}%
}  \nonumber \\
&=&\sum_{k=N-p+1}^{N}\eta _{k-\left( N-p\right) }2^{-k}+\frac{2^{N}-2^{p}+1}{%
2^{N+1}}.  \nonumber
\end{eqnarray}

\noindent (iii) $m=2\left( \mathrm{mod} 4\right) $

\noindent From the above lemma, we have

\[
\left| \left( A^{m}\right) _{\xi _{p+l}j_{l}}\right| ^{2}=\left\{ 
\begin{array}{ll}
2^{m} & \mbox{if }j_{l}\neq \xi _{p+l} \\ 
0 & \mbox{if }j_{l}=j_{p+l}
\end{array}
\right. ,\;\left| \left( A^{m+1}\right) _{\xi _{k-\left( N-p\right)
j_{k}}}\right| ^{2}=2^{m} 
\]

\noindent for any $l=1,\cdots ,N-p$ and $k=N-p+1,\cdots ,N$. Using this
formula the product of absolute squares can be expressed as

\begin{eqnarray*}
&&\prod\limits_{l=1}^{N-p}\left| \left( A^{m}\right) _{\xi
_{p+l}j_{l}}\right| ^{2}\prod\limits_{k=N-p+1}^{N}\left| \left(
A^{m+1}\right) _{\xi _{k-\left( N-p\right)}j_{k}}\right| \\
&=&\left\{ 
\begin{array}{ll}
\left( 2^{m}\right) ^{N-p}\left( 2^{m}\right) ^{p} & \mbox{if }j_{l}\neq \xi
_{p+l}\mbox{ for all }l=1,\cdots ,N-p \\ 
0 & \mbox{otherwise}
\end{array}
\right. \\
&=&\left\{ 
\begin{array}{ll}
2^{mN} & \mbox{if }j_{l}\neq \xi _{p+l}\mbox{ for all }l=1,\cdots ,N-p \\ 
0 & \mbox{otherwise}
\end{array}
\right. .
\end{eqnarray*}

\noindent Let $\eta _{p+l}=\xi _{p+l}+1\left( \mathrm{mod} 2\right)
,l=1,\cdots ,N-p$. It follows that

\begin{eqnarray*}
r_{n}^{\left( N\right) } &=&\frac{1}{2^{\left( m+1\right) N+p}}
\sum_{j_{1},\cdots ,j_{N}}\left\{ \left(
\sum_{k=1}^{N-p}j_{k}2^{N-k}\right)+\left(
\sum_{k=N-p+1}^{N}j_{k}2^{N-k}\right) +\frac{1}{2}\right\} \\
&&\times \prod\limits_{l=1}^{N-p}\left| \left( A^{m}\right)
_{\xi_{p+l}j_{l}}\right| ^{2}\prod\limits_{k=N-p+1}^{N}\left| \left(
A^{m+1}\right) _{\xi _{k-\left( N-p\right) }j_{k}}\right| ^{2} \\
&=&\frac{2^{mN}}{2^{\left( m+1\right) N+p}}\sum_{j_{N-p+1},\cdots
,j_{N}}\left\{ \left( \sum_{k=1}^{N-p}\eta _{p+k}2^{N-k}\right) +\left(
\sum_{k=N-p+1}^{N}j_{k}2^{N-k}\right) +\frac{1}{2}\right\}
\end{eqnarray*}

\noindent Substituting $\eta _{p+k}$ for $\xi _{p+k}$ in (\ref{5.13}), we
get 
\[
r_{n}^{\left( N\right) }=\sum_{k=1}^{N-p}\eta _{p+k}2^{-k}+\frac{2^{p}}{%
2^{N+1}}. 
\]

\noindent (iv) $m=3\left( \mathrm{mod} 4\right) $

\noindent From the above lemma, we have

\[
\left| \left( A^{m}\right) _{\xi _{p+l}j_{l}}\right| ^{2}=2^{m-1}, \left|
\left( A^{m+1}\right) _{\xi _{k-\left( N-p\right) j_{k}}}\right|^{2}=\left\{ 
\begin{array}{ll}
2^{m+1} & \mbox{if }j_{k}=\xi _{k-\left( N-p\right) } \\ 
0 & \mbox{if }j_{k}\neq \xi _{k-\left( N-p\right) }
\end{array}
\right. 
\]

\noindent for any $l=1,\cdots ,N-p$ and $k=N-p+1,\cdots ,N$. Using this
formula the product of absolute squares can be expressed as

\begin{eqnarray*}
&&\prod\limits_{l=1}^{N-p}\left| \left( A^{m}\right)
_{\xi_{p+l}j_{l}}\right| ^{2}\prod\limits_{k=N-p+1}^{N}\left| \left(
A^{m+1}\right) _{\xi _{k-\left( N-p\right)}j_{k}}\right| \\
&=&\left\{ 
\begin{array}{ll}
\left( 2^{m-1}\right) ^{N-p}\left( 2^{m+1}\right) ^{p} & \mbox{if }j=\xi
_{k-\left( N-p\right) }\mbox{ for all }k=N-p+1,\cdots ,N \\ 
0 & \mbox{otherwise}
\end{array}
\right. \\
&=&\left\{ 
\begin{array}{ll}
2^{\left( m-1\right) N+2p} & \mbox{if }j_{k}\neq \xi _{k-\left( N-p\right) }%
\mbox{ for all }k=N-p+1,\cdots ,N \\ 
0 & \mbox{otherwise}
\end{array}
\right. .
\end{eqnarray*}

\noindent \noindent (\ref{5.11}) can be rewritten as

\begin{eqnarray*}
r_{n}^{\left( N\right) } &=&\frac{1}{2^{\left( m+1\right) N+p}}
\sum_{j_{1},\cdots ,j_{N}}\left\{ \left(
\sum_{k=1}^{N-p}j_{k}2^{N-k}\right)+\left(
\sum_{k=N-p+1}^{N}j_{k}2^{N-k}\right) +\frac{1}{2}\right\} \\
&&\times \prod\limits_{l=1}^{N-p}\left| \left( A^{m}\right)
_{\xi_{p+l}j_{l}}\right| ^{2}\prod\limits_{k=N-p+1}^{N}\left|
\left(A^{m+1}\right) _{\xi _{k-\left( N-p\right) }j_{k}}\right| ^{2} \\
&=&\frac{2^{\left( m-1\right) N+2p}}{2^{\left( m+1\right) N+p}}%
\sum_{j_{N-p+1},\cdots ,j_{N}}\left\{ \left(
\sum_{k=1}^{N-p}j_{k}2^{N-k}\right) +\left( \sum_{k=N-p+1}^{N}\xi
_{k-\left(N-p\right) }2^{N-k}\right) +\frac{1}{2}\right\}
\end{eqnarray*}

\noindent Substituting $\xi _{k-\left( N-p\right) }$ for $\eta _{k-\left(
N-p\right) }$ in (\ref{5.13}), we get

\[
r_{n}^{\left( N\right) }=\sum_{k=N-p-1}^{N-p}\xi _{k-\left( N-p\right)
}2^{-k}+\frac{2^{N}-2^{p}+1}{2^{N+1}}. 
\]

\hspace{15cm}$\blacksquare $

\noindent \textbf{Proof of THEOREM 5.4:} For any $n=mN,m\in \mathbf{N}$,

\[
r_{n}^{\left( N\right) }=\frac{1}{2^{n}}\sum_{j=0}^{2^{N}-1}\frac{j+1/2}{
2^{N}}\prod\limits_{k=1}^{N}\left| \left( A^{m}\right) _{\xi
_{k}j_{k}}\right| ^{2}. 
\]

By a direct calculation, we obtain

\begin{eqnarray*}
r_{n}^{\left( N\right) } &=&\frac{1}{2^{n}}\sum_{j=0}^{2^{N}-1}\frac{j+1/2}{%
2^{N}}\prod\limits_{k=1}^{N}\left| \left( A^{m}\right)
_{\xi_{k}j_{k}}\right| ^{2} \\
&=&\frac{1}{2^{n+N}}\sum_{j=0}^{2^{N}-1}\left( j+1/2\right)
\prod\limits_{k=1}^{N}\left| \left( A^{m}\right) _{\xi _{k}j_{k}}\right| ^{2}
\\
&=&\frac{1}{2^{\left( m+1\right) N}}\sum_{j_{1},\cdots ,j_{N}}\left\{
\left(\sum_{k=1}^{N}j_{k}2^{N-k}\right) +\frac{1}{2}\right\}
\prod\limits_{k=1}^{N}\left| \left( A^{m}\right) _{\xi _{k}j_{k}}\right|^{2}
\end{eqnarray*}

\noindent (i) $m=0$ $\left( \mathrm{mod} 4\right) $

\noindent From the above lemma, we have

\[
\left| \left( A^{m}\right) _{\xi _{k}j_{k}}\right| ^{2}=\left\{ 
\begin{array}{ll}
2^{m} & \mbox{if }j_{k}=\xi _{k} \\ 
0 & \mbox{if }j_{k}\neq \xi _{k}
\end{array}
\right. 
\]

\noindent for any $k=1,\cdots ,N$ . Using this formula the product of
absolute squares can be expressed as

\[
\prod\limits_{l=1}^{N}\left| \left( A^{m}\right) _{\xi
_{k}j_{k}}\right|^{2}=\left\{ 
\begin{array}{ll}
2^{mN} & \mbox{if }j_{k}=\xi _{k}\mbox{ for all }k=1,\cdots ,N \\ 
0 & \mbox{otherwise}
\end{array}
\right. . 
\]

\noindent Using this formula the mean value $r_{n}^{\left( N\right) }$ of
the position operator can be expressed as

\begin{eqnarray*}
r_{n}^{\left( N\right) } &=&\frac{1}{2^{\left( m+1\right) N}}
\sum_{j_{1},\cdots ,j_{N}}\left\{ \left( \sum_{k=1}^{N}j_{k}2^{N-k}\right) +%
\frac{1}{2}\right\} \prod\limits_{k=1}^{N}\left| \left( A^{m}\right)
_{\xi_{k}j_{k}}\right| ^{2} \\
&=&\frac{2^{mN}}{2^{\left( m+1\right) N}}\left\{ \left( \sum_{k=1}^{N}\xi
_{k}2^{N-k}\right) +\frac{1}{2}\right\} \\
&=&\sum_{k=1}^{N}\xi _{k}2^{-k}+\frac{1}{2^{N+1}}.
\end{eqnarray*}

\noindent (ii) $m=1,3\left( \mathrm{mod} 4\right) $

\noindent From the above lemma, we have

\[
\left| \left( A^{m}\right) _{\xi _{k}j_{k}}\right| ^{2}=2^{m-1} 
\]

\noindent for any $k=1,\cdots ,N$ . Note that

\[
\prod\limits_{l=1}^{N}\left| \left( A^{m}\right) _{\xi
_{k}j_{k}}\right|^{2}=2^{\left( m-1\right) N}. 
\]

\noindent Using this formula the mean value $r_{n}^{\left( N\right) }$of the
position operator can be expressed as

\begin{eqnarray*}
r_{n}^{\left( N\right) } &=&\frac{1}{2^{\left( m+1\right) N}}
\sum_{j_{1},\cdots ,j_{N}}\left\{ \left( \sum_{k=1}^{N}j_{k}2^{N-k}\right) +%
\frac{1}{2}\right\} \prod\limits_{k=1}^{N}\left| \left( A^{m}\right)
_{\xi_{k}j_{k}}\right| ^{2} \\
&=&\frac{2^{\left( m-1\right) N}}{2^{\left( m+1\right) N}}
\sum_{j_{1},\cdots,j_{N}}\left\{ \left( \sum_{k=1}^{N}j_{k}2^{N-k}\right) +%
\frac{1}{2}\right\} \\
&=&\frac{1}{2^{2N}}.\left\{ \sum_{j_{1},\cdots
,j_{N}}\left(\sum_{k=1}^{N}j_{k}2^{N-k}\right) +\frac{1}{2}\left(
\sum_{j_{1},\cdots,j_{N}}1\right) \right\} \\
&=&\frac{1}{2^{2N}}\sum_{k=0}^{2^{N}-1}k+\frac{1}{2^{N+1}} \\
&=&\frac{1}{2^{2N}}2^{N-1}\left( 2^{N}-1\right) +\frac{1}{2^{N+1}} \\
&=&\frac{1}{2}.
\end{eqnarray*}

\noindent (iii) $m=2\left( \mathrm{mod} 4\right) $

\noindent From the above lemma, we have

\[
\left| \left( A^{m}\right) _{\xi _{k}j_{k}}\right| ^{2}=\left\{ 
\begin{array}{ll}
2^{m} & \mbox{if }j_{k}\neq \xi _{k} \\ 
0 & \mbox{if }j_{k}=\xi _{k}
\end{array}
\right. 
\]

\noindent for any $k=1,\cdots ,N$ . Using this formula the product of
absolute squares can be expressed as

\[
\prod\limits_{l=1}^{N}\left| \left( A^{m}\right) _{\xi
_{k}j_{k}}\right|^{2}=\left\{ 
\begin{array}{ll}
2^{mN} & \mbox{if }j_{k}\neq \xi _{k}\mbox{ for all }k=1,\cdots ,N \\ 
0 & \mbox{otherwise}
\end{array}
\right. . 
\]

\noindent Let $\eta _{k}=\xi _{k}+1\left( \mathrm{mod} 2\right)
,k=1,\cdots,N $. It follows that

\begin{eqnarray*}
r_{n}^{\left( N\right) } &=&\frac{1}{2^{\left( m+1\right) N}}%
\sum_{j_{1},\cdots ,j_{N}}\left\{ \left( \sum_{k=1}^{N}j_{k}2^{N-k}\right) +%
\frac{1}{2}\right\} \prod\limits_{k=1}^{N}\left| \left( A^{m}\right)
_{\xi_{k}j_{k}}\right| ^{2} \\
&=&\frac{2^{mN}}{2^{\left( m+1\right) N}}\left\{ \left(
\sum_{k=1}^{N}\eta_{k}2^{N-k}\right) +\frac{1}{2}\right\} \\
&=&\sum_{k=1}^{N}\eta _{k}2^{-k}+\frac{1}{2^{N+1}}\quad \blacksquare
\end{eqnarray*}

\end{document}